\documentclass[twocolumn,showpacs,floats,floatfix,superscriptaddress,aps,pra]{revtex4}
\usepackage{amssymb}
\usepackage{amsmath}
\usepackage{calc}
\usepackage{graphicx}
\usepackage{bm}

\begin{document}

\author{E. S. Kioseva}
\affiliation{Department of Physics, Sofia University, James Bourchier 5 blvd, 1164 Sofia,
Bulgaria}
\author{N. V. Vitanov}
\affiliation{Department of Physics, Sofia University, James Bourchier 5 blvd, 1164 Sofia,
Bulgaria}
\affiliation{Institute of Solid State Physics, Bulgarian Academy of Sciences,
Tsarigradsko chauss\'{e}e 72, 1784 Sofia, Bulgaria}
\title{Resonant excitation amidst dephasing: An exact analytic solution}
\date{\today }

\begin{abstract}
An exact analytic solution is presented for coherent resonant excitation of
a two-state quantum system driven by a time-dependent pulsed external field
with a hyperbolic-secant shape in the presence of dephasing. Analytic
results are derived for the amplitude and the phase shift of the damped Rabi
oscillations.
\end{abstract}

\maketitle


\section{Introduction\label{Sec-intro}}

Coherent resonant excitation represents an important notion in quantum
mechanics \cite{Shore,Allen-Eberly}. Resonant pulses of specific pulse areas
are widely used in a variety of fields in quantum physics, including nuclear
magnetic resonance, coherent atomic excitation, quantum information, and
others. Resonant excitation allows to establish a full control over the
quantum system, particularly in a two-state system, and realize any unitary
transformation (qubit rotation) in it. Particularly important
transformations include complete population transfer (transition probability 
$P=1$), complete population return ($P=0$) and Hadamard transformation ($P=%
\frac{1}{2}$).

Crucial conditions for resonant coherent excitation are resonance (the
frequency of the external field must be equal to the Bohr transition
frequency) and coherence. Deviations from resonance (detuning) are
detrimental and lead to rapid departure of the transition probability from
the desired value. In this respect, an alternative to resonant excitation is
provided by adiabatic excitation \cite{adiabatic}, which is robust against
such variations.

Even more crucial for resonant excitation is coherence. Incoherent
excitation, as described by Einstein's rate equations, allows only partial
population transfer, e.g., at most 50\% in a two-level system with equal
degeneracies of the two levels \cite{Shore}. Deviations from perfect
coherence, which can be described by the quantum Liouville or Bloch
equations, inevitably cause departure from the desired unitary
transformation. Two general types of decoherence processes can be present:
depopulation (e.g., due to spontaneous emission or ionisation) and dephasing
(e.g., due to elastic collisions, field fluctuations, coupling to the
environment, etc.).

In this paper, we present an exact analytic solution for pulsed resonant
excitation of a two-state system in the presence of pure dephasing.
Dephasing is recognised as one of the main obstacles in quantum information
and the availability of precise analytic estimates of its effect can be very
useful and important. We derive the exact solution of the Bloch equation for
a hyperbolic-secant pulse shape and a constant dephasing rate. We provide
examples of the general solution, which is expressed in terms of gamma
functions, in various special cases of interest, e.g. for pulses with
specific pulse areas. These results allow us to determine explicitly the
deviations from the desired probabilities caused by the dephasing. 

This paper is organised as follows. In Sec. \ref{Sec-model} we describe the
model and present the exact analytic solution, which is derived in the
appendix. In Sec. \ref{Sec-special} we analyse in some detail various
special cases. In Sec. \ref{Sec-conclusions} we give a summary of the
results.

\section{Analytic model\label{Sec-model}}

\subsection{The Bloch equation}

Dephasing processes can be incorporated into the quantum-mechanical
description of resonant excitation by including a phenomenological dephasing
rate $\Gamma $ into the Bloch equation \cite{Shore,Allen-Eberly},%
\begin{equation}
\frac{d}{dt}\left[ 
\begin{array}{c}
u(t) \\ 
v(t) \\ 
w(t)%
\end{array}%
\right] =\left[ 
\begin{array}{ccc}
-\Gamma  & -\Delta (t) & 0 \\ 
\Delta (t) & -\Gamma  & -\Omega (t) \\ 
0 & \Omega (t) & 0%
\end{array}%
\right] \left[ 
\begin{array}{c}
u(t) \\ 
v(t) \\ 
w(t)%
\end{array}%
\right] .  \label{Bloch equation}
\end{equation}%
This dephasing rate is the inverse of the transverse relaxation time $T_{2}$%
, $\Gamma =1/T_{2}$ \cite{Shore,Allen-Eberly}. Here $u(t)=2\Re \rho _{12}(t)$
and $v(t)=2\Im \rho _{12}(t)$ are the coherences\ and $w(t)=\rho
_{22}(t)-\rho _{11}(t)$ is the population inversion between the two states $%
\left\vert 1\right\rangle $ and $\left\vert 2\right\rangle $, with $\rho
_{mn}$ ($m,n=1,2$) being the density matrix elements. To be specific, we shall use the
language of laser-atom interactions, although the results apply to any
two-state system. In atomic excitation, the Rabi frequency $\Omega (t)$
quantifies the time-dependent dipole interaction between the two states and
it is proportional to the temporal envelope of the laser electric field $%
\mathbf{E}(t)$ and the transition dipole moment $\mathbf{d}$, $\hbar \Omega
(t)=-\mathbf{d}\cdot \mathbf{E}(t)$. The detuning $\Delta $ is the offset
between the transition frequency of the two-state system $\omega _{0}$ and
the laser field carrier frequency $\omega $, $\Delta =\omega _{0}-\omega $.

We shall solve the Bloch equation (\ref{Bloch equation}) with the initial
conditions%
\begin{equation}
u(-\infty )=v(-\infty )=0,\qquad w(-\infty )=-1,  \label{initial conditions}
\end{equation}%
which correspond to a system initially in state $\left\vert 1\right\rangle $%
: $\rho _{11}(-\infty )=1$, $\rho _{22}(-\infty )=0$. Our objective is to
find the Bloch vector $[u,v,w]^{T}$ as $t\rightarrow +\infty $, and
particularly, the population inversion $w(\infty )$, since the coherences
vanish at infinity due to the dephasing.

\subsection{The model}

We suppose an exact resonance, a hyperbolic-secant pulse and a constant
dephasing rate, 
\begin{subequations}
\label{model}
\begin{eqnarray}
\Delta (t) &=&0,  \label{model-Delta} \\
\Omega (t) &=&\Omega _{0}\text{sech}(t/T),  \label{model-Omega} \\
\Gamma (t) &=&\text{const}.  \label{model-Gamma}
\end{eqnarray}%
The dephasing rate $\Gamma $ is a positive constant and $T$ is the
characteristic pulse width. The peak Rabi frequency $\Omega _{0}$ will be
assumed also positive without loss of generality. The pulse area of the sech
pulse (\ref{model-Omega}) is 
\end{subequations}
\begin{equation}
\mathcal{A}=\int_{-\infty }^{+\infty }\Omega (t)dt=\pi \Omega _{0}T.
\label{area}
\end{equation}

\subsection{The exact solution in the coherent limit}

For $\Gamma =0$, the Bloch equation (\ref{Bloch equation}) is solved exactly 
\cite{Shore,Allen-Eberly}, 
\begin{equation}
w(\infty )=-\cos \mathcal{A}.  \label{w-resonance}
\end{equation}

Of particular interest are the cases when $\mathcal{A}$\ is equal to an
integer or half-integer multiple of $\pi $. There are three cases of special
significance.

\begin{itemize}
\item The \emph{odd-}$\pi $\emph{\ pulses} with area%
\begin{equation}
\mathcal{A}=(2n+1)\pi ,\qquad (n=0,1,2,\ldots ),  \label{odd pi}
\end{equation}%
invert the population, $w(\infty )=1$ ($\rho _{11}=0,\rho _{22}=1$). A
special case is the $\pi $ pulse with $\mathcal{A}=\pi $.

\item The \emph{even-}$\pi $\emph{\ pulses} with area%
\begin{equation}
\mathcal{A}=2n\pi ,\qquad (n=0,1,2,\ldots ),  \label{even pi}
\end{equation}%
restore the population to the initial state, $w(\infty )=-1$ ($\rho
_{11}=1,\rho _{22}=0$). A special case is the $2\pi $-pulse with $\mathcal{A}%
=2\pi $.

\item The \emph{half-integer-}$\pi $\emph{\ pulses} with area%
\begin{equation}
\mathcal{A}=(2n+1)\frac{\pi }{2},\qquad (n=0,1,2,\ldots ),  \label{half pi}
\end{equation}%
create an equal superposition between states 1 and 2, $w(\infty )=0$ ($\rho
_{11}=\rho _{22}=\frac{1}{2}$). A special case is the half-$\pi $ pulse with 
$\mathcal{A}=\pi /2$.
\end{itemize}

All these three cases are of great importance and such pulses are widely
used in various applications in quantum physics, e.g. in nuclear magnetic
resonance, coherent atomic excitation and quantum information. We shall
therefore pay special attention to these cases in the analytic solution,
which we shall derive below.

\subsection{The exact solution with dephasing}

Because of the resonance condition (\ref{model-Delta}), it follows from Eq. (%
\ref{Bloch equation}) that the equation for $\dot{u}$ decouples (with the
overdot denoting a time derivative),%
\begin{equation}
\dot{u}(t)=-\Gamma u(t),  \label{du(t)}
\end{equation}%
and can be solved independently,%
\begin{equation}
u(t)=u(-\infty )e^{-\Gamma t}=0,  \label{u(t)}
\end{equation}%
where we have used the initial conditions (\ref{initial conditions}).

We change the Bloch variable $v(t)=-ix(t)$ and Eq. (\ref{Bloch equation}) is
reduced to two coupled equations, 
\begin{subequations}
\label{Bloch equation reduced}
\begin{eqnarray}
i\dot{x}(t) &=&-i\Gamma x(t)+\Omega (t)w(t),  \label{dx(t)} \\
i\dot{w}(t) &=&\Omega (t)x(t).  \label{dw(t)}
\end{eqnarray}%
These equations resemble the Schr\"{o}dinger equation for the Rosen-Zener
model \cite{RZ} with irreversible loss from one of the states \cite%
{Vitanov97,Vitanov98}. For reader's convenience, the derivation is adopted
to our case and given in the Appendix.

The exact solution for $w(t)$ as $t\rightarrow +\infty $ reads 
\end{subequations}
\begin{equation}
w(\infty )=-\frac{\Gamma ^{2}\left( \frac{1}{2}+\gamma \right) }{\Gamma
\left( \frac{1}{2}+\gamma +\alpha \right) \Gamma \left( \frac{1}{2}+\gamma
-\alpha \right) },  \label{w(infinity)}
\end{equation}%
where $\Gamma (z)$ is the gamma function \cite{Erdelyi,AS,GR} and the
dimensionless parameters $\alpha $ and $\gamma $ are defined as%
\begin{equation}
\alpha =\Omega _{0}T,\qquad \gamma =\frac{\Gamma T}{2}.  \label{alpha, gamma}
\end{equation}%
Because of the dephasing, the coherences vanish,%
\begin{equation}
u(\infty )=v(\infty )=0.  \label{u,v}
\end{equation}

By using the reflection formula $\Gamma (z)\Gamma (1-z)=\pi /\sin (\pi z)$ 
\cite{AS}, Eq. (\ref{w(infinity)}) can be written also as%
\begin{equation}
w(\infty )=-\frac{\Gamma ^{2}\left( \frac{1}{2}+\gamma \right) \Gamma \left( 
\frac{1}{2}-\gamma +\alpha \right) }{\pi \Gamma \left( \frac{1}{2}+\gamma
+\alpha \right) }\cos \pi \left( \alpha -\gamma \right) .  \label{w-cos}
\end{equation}%
For $\gamma =0$, Eq. (\ref{w-cos}) reduces to the lossless solution (\ref%
{w-resonance}). Equation (\ref{w-cos}) shows that $w(\infty )$ vanishes
whenever the cosine factor vanishes, i.e. for $\left\vert \alpha -\gamma
\right\vert =\frac{1}{2},\frac{3}{2},\frac{5}{2},\cdots $. Hence the values
of the pulse area, for which an equal superposition between states $%
\left\vert 1\right\rangle $ and $\left\vert 2\right\rangle $ is created ($%
\rho _{11}=\rho _{22}=\frac{1}{2}$), are shifted from their half-$\pi $
values (\ref{half pi}),%
\begin{equation}
\mathcal{A}=\left( 2n+1+\Gamma T\right) \frac{\pi }{2}.
\label{half-pi dephasing}
\end{equation}

Several approximations to $w(\infty )$ are given below in some special cases.

\section{Special cases\label{Sec-special}}

\subsection{Specific pulse areas}

For $\alpha =n$ ($n\pi $ pulse), where $n$ is an integer, Eq. (\ref%
{w(infinity)}) reduces to%
\begin{equation}
w(\infty )=-\prod_{k=0}^{n-1}\frac{2\gamma -1-2k}{2\gamma +1+2k},
\label{w-integer}
\end{equation}%
where we have used repeatedly the recurrence relation $\Gamma (z+1)=z\Gamma
(z)$. It follows from Eq. (\ref{w-integer}) that $w(\infty )=0$ for $\gamma =%
\frac{1}{2}$ for any integer $\alpha =n\geqq 1$.

For $\alpha =1$ ($\pi $ pulse) we have%
\begin{equation}
w(\infty )=\frac{1-2\gamma }{1+2\gamma }.  \label{w-pi}
\end{equation}%
Hence the population inversion is a decreasing function of $\gamma $, which
decreases to $w_{\varepsilon }=1-\varepsilon $ for%
\begin{equation}
\gamma _{\varepsilon }=\frac{1}{2}\frac{1-w_{\varepsilon }}{1+w_{\varepsilon
}}.
\end{equation}%
For $w_{\varepsilon }=0.9$, $0.5$ and $0$ we find $\gamma _{\varepsilon }=%
\frac{1}{38}$, $\frac{1}{6}$ and $\frac{1}{2}$, i.e. the inversion decreases
very rapidly as $\gamma $ increases.

Similar simple expressions can be derived from Eq. (\ref{w-integer}) for
other cases of integer-$\pi $ pulses. 

For $\alpha =n+\frac{1}{2}$ (half-integer-$\pi $ pulse) we find from Eq. (%
\ref{w(infinity)}) that%
\begin{equation}
w(\infty )=-\frac{\gamma \Gamma ^{2}\left( \frac{1}{2}+\gamma \right) }{%
\Gamma ^{2}\left( 1+\gamma \right) }\prod_{k=1}^{n}\frac{\gamma -k}{\gamma +k%
},  \label{w-half-integer}
\end{equation}%
The factor in front of the product gives $w(\infty )$ for $\alpha =\frac{1}{2%
}$ ($\pi /2$ pulse).


\begin{figure}[tb]
\includegraphics[width=80mm]{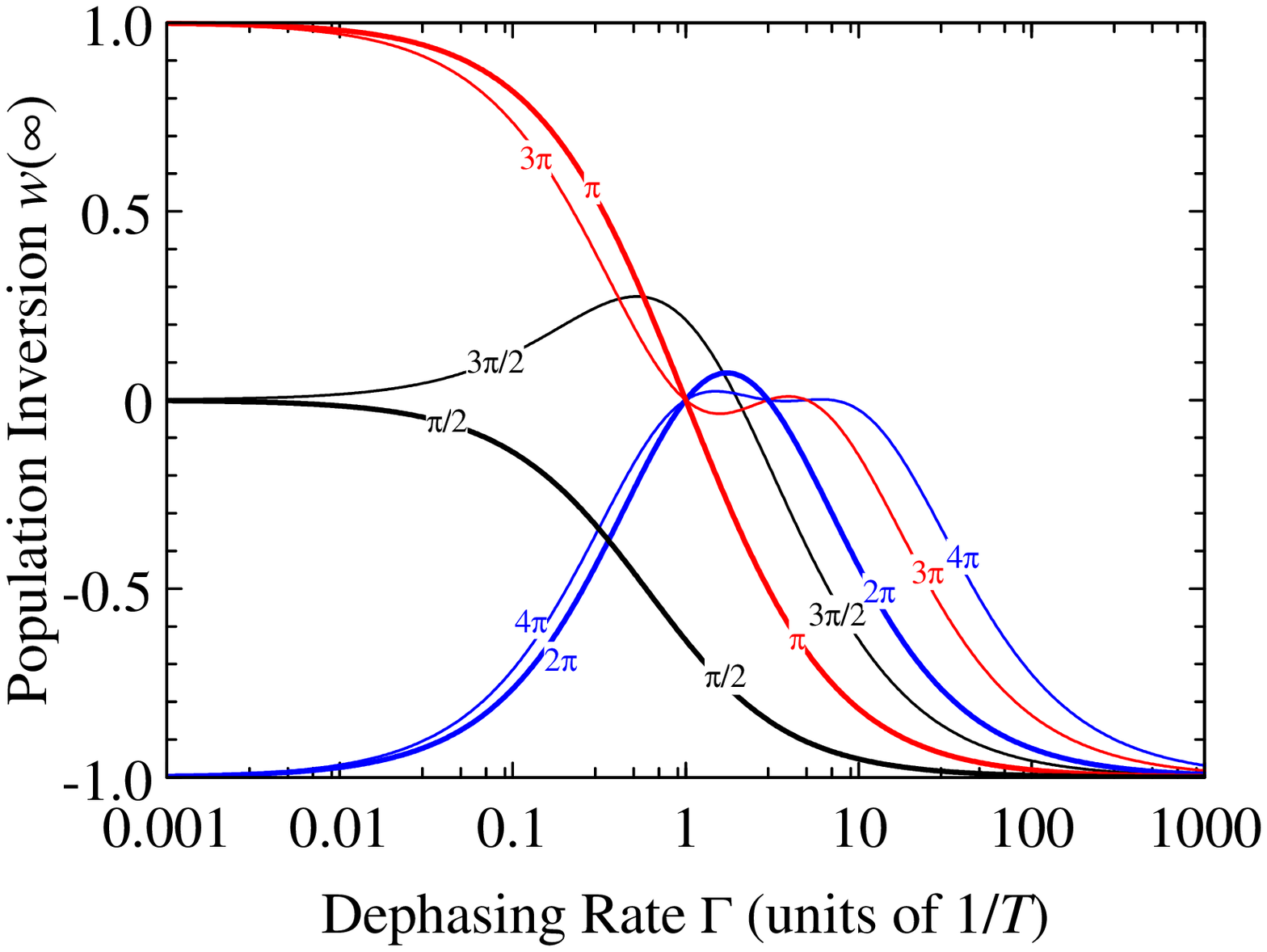}
\caption{The population inversion, Eq. (\protect\ref{w-integer}) for integer-%
$\protect\pi $ pulses and Eq. (\protect\ref{w-half-integer}) for
half-integer-$\protect\pi $ pulses, against the dephasing rate for different
values of the pulse area $\mathcal{A}=\protect\pi \protect\alpha =\protect%
\pi \Omega _{0}T$, denoted on the respective curve.}
\label{Fig-gamma}
\end{figure}

In Fig. \ref{Fig-gamma} the population inversion $w(\infty )$ is plotted
against the dephasing rate for different pulse areas.

\subsection{Weak dephasing}

When $\gamma \ll 1$ we use the relation $\Gamma (a+\gamma )=\Gamma (a)\left[
1+\psi (a)\gamma +\mathcal{O}(\gamma ^{2})\right]$, where $\psi (a)$ is the
psi function, and the relation $\psi (1/2)=-\left( c+2\ln
2\right)$, where $c=0.5772156649\ldots $ is the Euler-Mascheroni
constant \cite{AS,Erdelyi}. We thus find from Eq. (\ref{w-cos}) that%
\begin{eqnarray}
w(\infty ) &\approx &-\left\{ 1-2\left[ c+2\ln 2+\psi \left( \frac{1}{2}%
+\alpha \right) \right] \gamma +\mathcal{O}(\gamma ^{2})\right\}   \notag \\
&&\times \cos \left[ \pi \left( \alpha -\gamma \right) \right] .
\label{w weak dephasing}
\end{eqnarray}%
The first factor $\left\{ \cdots \right\} $ describes the amplitude of the
damped Rabi oscillations and the $\cos $ factor describes the phase of the
oscillations. The maxima and the minima of these oscillations are shifted by 
$\pi \gamma $ (if the small additional shift from the damped amplitude is
neglected) from their coherent values (\ref{odd pi}) and (\ref{even pi}),
respectively. The factor $\left\{ \cdots \right\} $ in Eq. (\ref{w weak
dephasing}) displays explicitly the damping of the amplitude and its
departure from 1 as $\gamma $ rises from zero. Since $\psi \left( \frac{1}{2}%
+\alpha \right) $ is an increasing function of $\alpha $ \cite{AS}, the
damping effect is stronger for larger pulse areas, which is shown explicitly
below.

For $\alpha =\gamma +n$, near the $n$th extremum, we find from Eq. (\ref{w
weak dephasing}) by using Eq. (6.3.4) of \cite{AS} for $\psi (n+\frac{1}{2})$
that%
\begin{equation}
w(\infty )\approx \left( -1\right) ^{n+1}\left[ 1-4\gamma \sum_{k=1}^{n}%
\frac{1}{2k-1}+\mathcal{O}(\gamma ^{2})\right] .  \label{w weak extrema}
\end{equation}%
For $n=1-4$, Eq. (\ref{w weak extrema}) gives 
\begin{subequations}
\label{extrema}
\begin{eqnarray}
n &=&1:\quad w(\infty )\approx 1-4\gamma +\mathcal{O}(\gamma ^{2}),
\label{n=1} \\
n &=&2:\quad w(\infty )\approx -1+\frac{16}{3}\gamma +\mathcal{O}(\gamma
^{2}),  \label{n=2} \\
n &=&3:\quad w(\infty )\approx 1-\frac{92}{15}\gamma +\mathcal{O}(\gamma
^{2}),  \label{n=3} \\
n &=&4:\quad w(\infty )\approx -1+\frac{704}{105}\gamma +\mathcal{O}(\gamma
^{2}).  \label{n=4}
\end{eqnarray}%
The cases of $n=1$ and 3 correspond to the first and second maxima ($\pi $
and $3\pi $ pulses), and $n=2$ and 4 to the first and second minima ($2\pi $
and $4\pi $ pulses). Equations (\ref{w weak extrema}) and (\ref{extrema})
show explicitly how the values of the population inversion for these $n\pi $
pulses depart from their values $\pm 1$ as $\gamma $ departs from zero.
These equations also demonstrate that the effect of dephasing is stronger
for larger pulse areas (since the coefficient in front of $\gamma $
increases with $n$), which is indeed seen in Fig. \ref{Fig-gamma}.


\begin{figure}[tb]
\includegraphics[width=80mm]{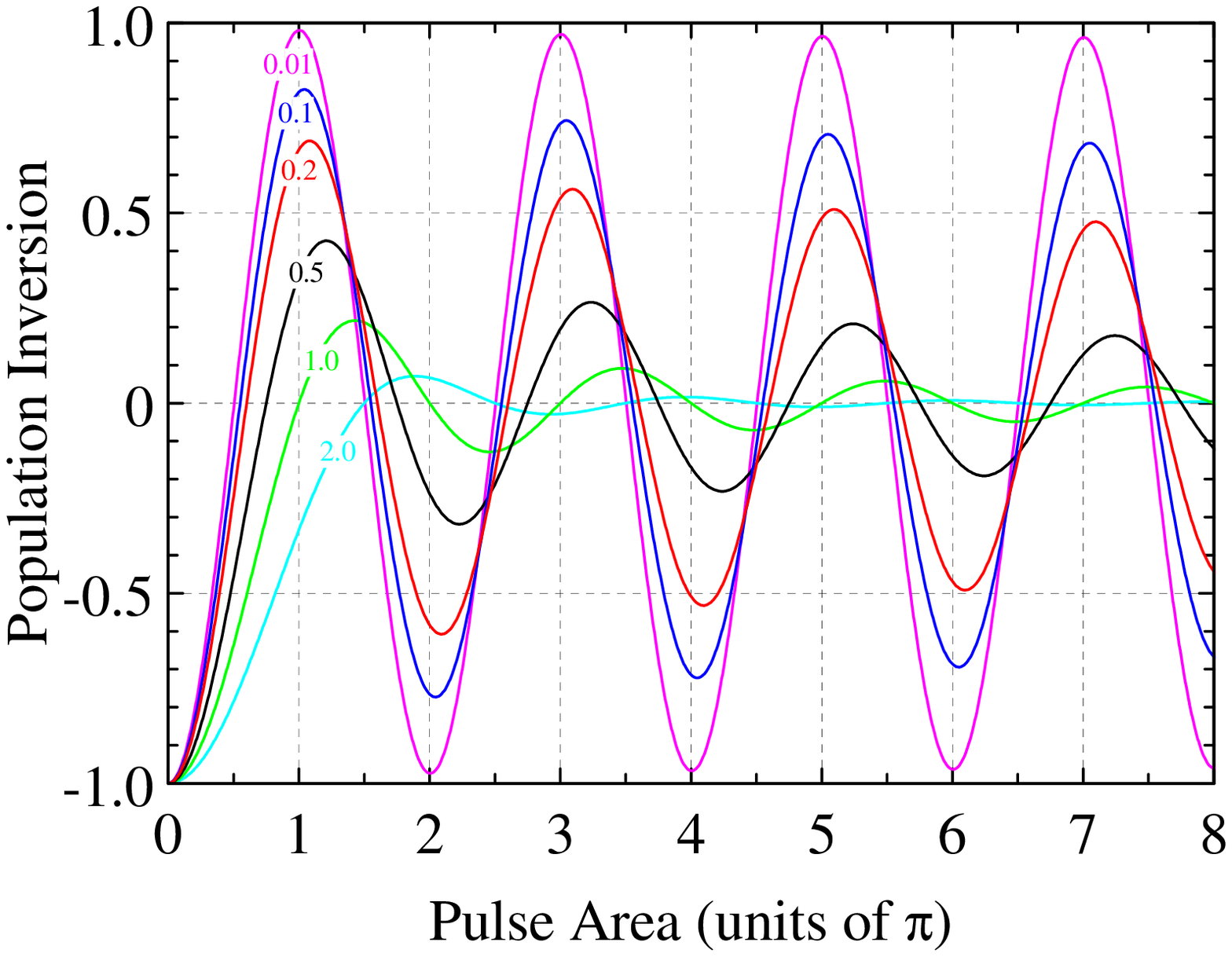}
\caption{The population inversion (\protect\ref{w(infinity)}) against the
pulse area $\mathcal{A}=\protect\pi \protect\alpha $ for different values of
the dephasing rate, $\Gamma T=0.01,0.1,$ $0.2,$ $0.5,1,2$ (denoted on the
respective curve).}
\label{Fig-alpha}
\end{figure}

In Fig. \ref{Fig-alpha} the population inversion is plotted against the
pulse area for several values of the dephasing rate. As predicted, for $%
\gamma =\frac{1}{2}$ ($\Gamma T=1$), the nodes of $w$ are situated at pulse
areas $\mathcal{A}=n\pi $, where in the absence of dephasing one finds the
extrema. For $\gamma =1$ ($\Gamma T=2$), the maxima are situated
approximately at $\alpha =2,4,6,\ldots $, where one finds the minima (even-$%
\pi $ pulses) for $\gamma =0$; likewise, the minima are situated
approximately at $\alpha =3,5,7,\ldots $, where one finds the maxima (odd-$%
\pi $ pulses) for $\gamma =0$.

\subsection{Strong dephasing}

For $\gamma \gg \alpha ,1$, the population inversion $w(\infty )$ has the
asymptotics 
\end{subequations}
\begin{equation}
w(\infty )\sim -\exp \left[ -\frac{\alpha ^{2}}{\gamma }+\mathcal{O}(\gamma
^{-3})\right] ,  \label{w strong dephasing}
\end{equation}%
which is obtained from Eq. (\ref{w(infinity)}) by using the Stirling
asymptotic expansion \cite{AS}, 
\begin{eqnarray}
\Gamma (z) &\sim &\sqrt{2\pi }z^{z-\frac{1}{2}}\exp \left[ -z+\frac{1}{12z}+%
\mathcal{O}\left( \left\vert z\right\vert ^{-3}\right) \right] ,
\label{Stirling} \\
&&\qquad \left( \left\vert \arg z\right\vert <\pi ,\left\vert z\right\vert
\gg 1\right) .  \notag
\end{eqnarray}
The inversion $w(\infty )$ decreases in a Gaussian fashion against $\alpha $%
. For large $\gamma $, $w$ tends to its initial value of $-1$, rather than
to the incoherent limit $w=0$, which is a result of quantum overdamping \cite%
{Vitanov97}. This behavior is indeed seen in Fig. \ref{Fig-gamma}, where we
have verified that Eq. (\ref{w strong dephasing}) describes very accurately
the asymptotic decrease of $w(\infty )$ for large $\gamma $ (not shown for
simplicity).

\subsection{Large pulse area}

When $\alpha \gg \gamma ,1$, the population inversion $w(\infty )$ has the
following behavior 
\begin{equation}
w(\infty )\sim -\frac{1}{\pi }\Gamma ^{2}\left( \frac{1}{2}+\gamma \right)
\alpha ^{-2\gamma }\left[ 1+\mathcal{O}(\alpha ^{-2})\right] \cos \pi
(\alpha -\gamma ),  \label{w large area}
\end{equation}%
which is derived from Eq. (\ref{w-cos}) by using Eq. (\ref{Stirling}).
Equation (\ref{w large area}) shows that as $\alpha $ increases, the
oscillation amplitude vanishes as $\alpha ^{-2\gamma }$ and for sufficiently
large pulse areas,%
\begin{equation}
\mathcal{A}\gtrsim \mathcal{A}_{\varepsilon }=\pi \left[ \frac{\pi
\varepsilon }{\Gamma ^{2}\left( \frac{1}{2}+\gamma \right) }\right] ^{-\frac{%
1}{2\gamma }},  \label{A limit}
\end{equation}%
the population inversion decreases below $\varepsilon $ ($\left\vert
w\right\vert \lesssim \varepsilon $), i.e. the two-state system evolves
towards a completely incoherent superposition of states $\left\vert
1\right\rangle $ and $\left\vert 2\right\rangle $ ($\rho _{11}=\rho _{22}=%
\frac{1}{2}$, $\rho _{12}=0$). For instance, $\mathcal{A}_{0.1}\approx
1.58\times 10^{9}\pi $, $415\pi $, and $3.18\pi $ for $\gamma =0.1$, $0.3$,
and $1$, respectively.

\section{Conclusions\label{Sec-conclusions}}

In this paper we have presented an exact analytic solution for resonant
excitation induced by a pulse with a hyperbolic-secant shape in the presence
of dephasing processes. The exact solution (\ref{w(infinity)}) is given in
terms of gamma functions. Dephasing affects the Rabi oscillations in two
ways: shifting the oscillation phase by approximately $\pi \Gamma T/2$ and
damping the oscillation amplitude: the larger the pulse area, the stronger
the damping. The implication is that one cannot reduce the dephasing-induced
losses of efficiency by increasing the intensity of the field (e.g.
replacing a $\pi $ pulse by a $3\pi $ pulse) since this will actually
increase the losses.

Various special cases of pulses with specific areas have been considered and
various limits have been derived in terms of elementary functions. The
results provide explicit and simple estimates of the effect of dephasing on
resonant excitation, e.g. in the cases of $\pi $, $2\pi $ and $\pi /2$
pulses, which are of great importance and widely used in many fields.

\subsection*{Acknowledgements}

This work has been supported by the EU Transfer of Knowledge project CAMEL,
the Alexander von Humboldt Foundation, and the EU Research and Training
Network QUACS (HPRN-CT-2002-00309).

\appendix

\section{Exact solution}

The first step in solving Eqs. (\ref{Bloch equation reduced}) is to decouple
them by differentiating the equation for $\dot{w}$ and replacing $v$ and $%
\dot{v}$, found from Eqs. (\ref{Bloch equation reduced}); this gives%
\begin{equation}
\ddot{w}-\left( \Gamma +\frac{\dot{\Omega}}{\Omega }\right) \dot{w}+\Omega
^{2}w=0,  \label{Bloch equation decoupled}
\end{equation}%
with an overdot denoting $d/dt$. We change the independent variable from $t$
to $z(t)=\frac{1}{2}[\tanh (t/T)+1]$; hence $z(-\infty )=0$ and $z(+\infty
)=1$. Then 
\begin{equation}
z(1-z)W^{\prime \prime }+\left( \frac{1}{2}+\gamma -z\right) W^{\prime
}+\alpha ^{2}W=0,  \label{Cb-eqn}
\end{equation}%
where $^{\prime }\equiv d/dz$, $W\left[ z(t)\right] =w(t)$ and $\alpha $ and 
$\gamma $ are defined by Eqs. (\ref{alpha, gamma}). This equation has the
same form as the Gauss hypergeometric equation \cite{AS,GR}, 
\begin{equation}
z(1-z)F^{\prime \prime }+\left[ \nu -(\lambda +\mu +1)z\right] F^{\prime
}-\lambda \mu F=0,  \label{Gauss-eqn}
\end{equation}%
upon the identification 
\begin{equation}
\lambda =\alpha ,\qquad \mu =-\alpha ,\qquad \nu =\frac{1}{2}+\gamma .
\label{parameters}
\end{equation}

The complete solution of this equation, expressed by a superposition of two
linearly independent solutions of Eq. (\ref{Cb-eqn}), depends upon the value
of $\nu $.

\emph{The case }$\nu \neq 1,2,3,\ldots $According to Sec. 9.153.1 of \cite%
{GR}, the solution of Eq. (\ref{Cb-eqn}) can be expressed in terms of the
Gauss hypergeometric function \cite{AS,GR} as 
\begin{eqnarray}
W(z) &=&A_{1}F(\lambda ,\mu ;\nu ;z)  \notag \\
&&+A_{2}z^{1-\nu }F(\lambda +1-\nu ,\mu +1-\nu ;2-\nu ;z).  \label{Cb-sol}
\end{eqnarray}%
From here and using Eq. (\ref{Bloch equation reduced}), it can be found that%
\begin{eqnarray}
V(z) &=&\frac{i\sqrt{z(1-z)}}{\alpha }\bigg[A_{1}\frac{\lambda \mu }{\nu }%
F(\lambda +1,\mu +1;\nu +1;z)  \notag \\
&+&A_{2}(1-\nu )z^{-\nu }F(\lambda +1-\nu ,\mu +1-\nu ;1-\nu ;z)\bigg],
\end{eqnarray}%
with $V[z(t)]=v(t)$, where Eqs. (15.2.1) and (15.2.4) of \cite{AS} have been
used. The integration constants $A_{1}$ and $A_{2}$ can be determined from
the initial conditions (\ref{initial conditions}),%
\begin{equation}
A_{1}=-1,\qquad A_{2}=0.  \label{initial conditions A1,A2}
\end{equation}%
Hence $w(\infty )=W(1)=-F(\lambda ,\mu ;\nu ;1)$ or%
\begin{equation}
w(\infty )=-\frac{\Gamma (\nu )\Gamma (\nu -\lambda -\mu )}{\Gamma (\nu
-\lambda )\Gamma (\nu -\mu )},  \label{a}
\end{equation}%
where Eq. (15.1.20) of \cite{AS} has been used. Referring to Eqs. (\ref%
{parameters}), one obtains Eq. (\ref{w(infinity)}).

Equation (\ref{a}) has been derived under the assumption that $\nu \neq
1,2,3,\ldots $; then the two terms in Eq. (\ref{Cb-sol}) are linearly
independent. Suppose now that $\nu =n$ where $n=1,2,3,\ldots $, that is $%
\gamma =\frac{1}{2},\frac{3}{2},\frac{5}{2},\ldots $; Then the two terms in
Eq. (\ref{Cb-sol}) are linearly dependent for $\nu =1$ while the second term
is not defined for $\nu =2,3,4,\ldots $

\emph{The case }$\nu =1$. According to Sec. 9.153.2 of \cite{GR}, the
solution of Eq. (\ref{Cb-eqn}) for $\nu =1$ is%
\begin{eqnarray}
W(z) &=&A_{1}F(\lambda ,\mu ;1;z)  \notag \\
&+&A_{2}\left[ F(\lambda ,\mu ;1;z)\ln z+\sum_{k=1}^{\infty }\frac{(\lambda
)_{k}(\mu )_{k}\psi _{1}}{(k!)^{2}}z^{k}\right] ,
\end{eqnarray}%
with $(x)_{k}=\Gamma (x+k)/\Gamma (x)$ and $\psi _{1}=\psi (\lambda +k)-\psi
(\lambda )+\psi (\mu +k)-\psi (\mu )-2\psi (k+1)+2\psi (1)$, $\psi (x)$
being the psi-function \cite{AS}. Since the second term diverges for $z=0$,
the initial conditions (\ref{initial conditions}) require Eqs. (\ref{initial
conditions A1,A2}) to be satisfied and Eq. (\ref{a}) applies again.

\emph{The case }$\nu =n+1$\emph{\ }$(n=1,2,3,\ldots )$\emph{\ and }$\lambda
,\mu \neq 0,1,2,\ldots ,n-1$\emph{. }According to Sec. 9.153.3 of \cite{GR}
and Eq. (15.5.19) of \cite{AS}, the solution of Eq. (\ref{Cb-eqn}) in this
case is%
\begin{eqnarray}
W(z) &=&A_{1}F(\lambda ,\mu ;n+1;z)+A_{2}\bigg[F(\lambda ,\mu ;n+1;z)\ln z 
\notag \\
&+&\sum_{k=1}^{\infty }\frac{(\lambda )_{k}(\mu )_{k}\psi _{n+1}}{(n+1)_{k}k!%
}z^{k}-\sum_{k=1}^{n}\frac{(k-1)!(-n)_{k}}{(1-\lambda )_{k}(1-\mu )_{k}}%
z^{-k}\bigg]
\end{eqnarray}%
with $\psi _{n+1}=\psi (\lambda +k)-\psi (\lambda )+\psi (\mu +k)-\psi (\mu
)-\psi (n+1+k)+\psi (n+1)-\psi (1+k)+\psi (1)$. Again, the second term
diverges for $z=0$ and the initial conditions (\ref{initial conditions})
require Eqs. (\ref{initial conditions A1,A2}) to be satisfied and hence, Eq.
(\ref{a}) holds again.

\emph{The case }$\nu =n+1$\emph{\ }$(n=1,2,3,\ldots )$\emph{\ and }$\lambda $%
\emph{\ or }$\mu =0,1,2,\ldots ,n-1$. Suppose that $\lambda =m+1<n+1$. Then,
according to Sec. 9.153.4 of \cite{GR}, the solution is given by Eq. (\ref%
{Cb-sol}), in which the second hypergeometric function reduces to a
polynomial in $z^{-1}$. Since it diverges for $z=0$, the initial conditions (%
\ref{initial conditions}) require Eqs. (\ref{initial conditions A1,A2}) to
be satisfied and Eq. (\ref{a}) holds again.

In conclusion, for the model (\ref{model}), in all cases the final
population inversion $w(+\infty )$ is given by Eq. (\ref{a}).

\end{document}